\documentclass[twocolumn,epsf,prb,amsmath,amssymb,showpacs]{revtex4}
\usepackage{bm}
\usepackage{graphicx}
\usepackage{amsmath}
\usepackage{amssymb}
\usepackage{sidecap}
\usepackage{dcolumn}
\usepackage{color}
\usepackage{epstopdf}
\usepackage{ifpdf}
\usepackage{sidecap}

\begin{document}
\title{Graphene Hall bar with an asymmetric pn-junction}

\author{S. P. Milovanovi\'{c}}\email{slavisa.milovanovic@gmail.com}
\affiliation{Departement Fysica, Universiteit Antwerpen \\
Groenenborgerlaan 171, B-2020 Antwerpen, Belgium}

\author{M.~Ramezani Masir}\email{mrmphys@gmail.com}
\affiliation{Departement Fysica, Universiteit Antwerpen \\
Groenenborgerlaan 171, B-2020 Antwerpen, Belgium}

\author{F.~M.~Peeters}\email{francois.peeters@ua.ac.be}
\affiliation{Departement Fysica, Universiteit Antwerpen \\
Groenenborgerlaan 171, B-2020 Antwerpen, Belgium}
\begin{abstract}
We investigated the magnetic field dependence of the Hall and the bend resistances in the ballistic regime for a single layer graphene Hall bar structure containing a pn-junction. When both regions are n-type the Hall resistance dominates and Hall type of plateaus are formed. These plateaus occur as a consequence of the restriction on the angle imposed by Snell's law allowing only electrons with a certain initial angles to transmit though the potential step. The size of the plateau and its position is determined by the position of the potential interface as well as the value of the applied potential. When the second region is p-type the bend resistance dominates which is asymmetric in field due to the presence of snake states. Changing the position of the pn-interface in the Hall bar strongly affects these states and therefore the bend resistance is also changed. Changing the applied potential we observe that the bend resistance exhibits a peak around the charge-neutrality point (CNP) which is independent of the position of the pn-interface, while the Hall resistance shows a sign reversal when the CNP is crossed, which is in very good agreement with a recent experiment [J. R. Williams \textit{et al.}, Phys. Rev. Lett. \textbf{107}, 046602(2011)].
\end{abstract}


\date{\today}

\maketitle

\section{Introduction}
\label{Int}
Graphene, a one-atom-thick monolayer of graphite with a honeycomb
lattice structure, is a new material which has risen tremendous
interest in recent years. Chiral massless particles with a linear
spectrum near the $K$ and $K'$ points \cite{f1,f2} cause perfect
transmission through arbitrarily high and wide barriers, referred to
as Klein tunneling \cite{f3,f4,f5,f7}. The unusual band structure of single-layer graphene makes it a good candidate for novel sensors and an intriguing material
for electronic devices. The meta-material
character of pn-structures in graphene was pointed out in
Ref. \onlinecite{f4}, and focusing of electronic waves was
proposed\cite{mog10,has10}.
Recently, a graphene pn-junction with a possibility of separate control of carrier density in both regions using a pair of gates was reported. The density in each region could be varied across the charge neutrality point, allowing pn-, pp-, and nn-junctions to be formed at the interface within a single sheet.
Moreover the presence of snake states
along the pn-interface was predicted\cite{Ma1,Ma2} and experiments
on such systems were undertaken
recently\cite{Marcus1,Marcus0}. The meta-material properties of the
above mentioned pn-structures resulted in the expectancy of
controlling the wave function of the electron, in particular, the width of
electron beams by means of a superlattice  known as collimation
\cite{Cheol}. Qualitatively, the meta-material properties of
the pn-junctions in graphene can be understood by inspecting classical
trajectories\cite{A2}, or using ray optics as it is called for the
case of electromagnetic phenomena\cite{A1}. A similar classical-type of
simulation was done recently\cite{A3} for a graphene Hall bar containing
a pn-junction in the center of the cross.
\begin{figure}[ht]
\begin{center}
\includegraphics[width=6cm]{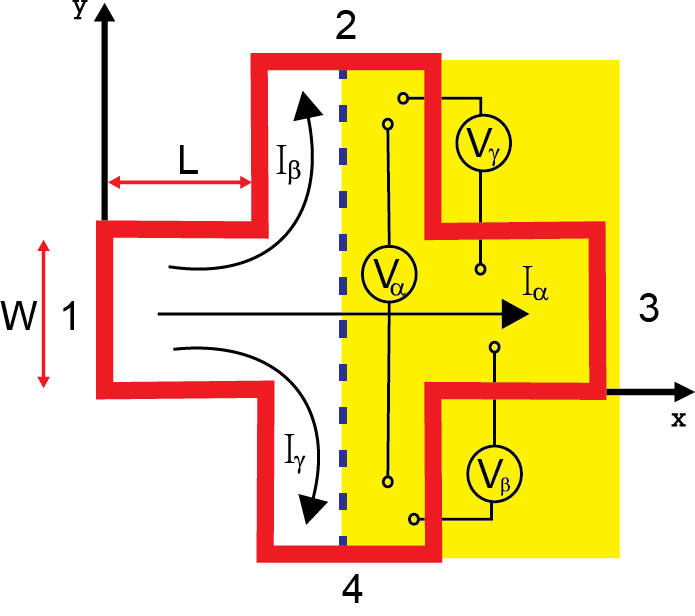}
\caption{(Color online) Schematics of the Hall bar with the pn-junction.}
\label{fig1}
\end{center}
\end{figure}

The Klein tunneling, Veselango lensing effect, individual control of carriers in two adjacent regions together with a linear energy-momentum relation makes a graphene Hall bar with a pn-junction a very interesting system. The possibility of fabricating such a device motivated us to examine the effect of the position of the pn-interface on the transport properties. We will examine what happens with the bend and the Hall resistance as we shift the pn-junction interface from the end of lead 1, positioned at $D = L$, to the beginning of lead 3, at $D = L+W$ (Fig. \ref{fig1}). To obtain those resistances we need to
calculate the different transmission, $T_{ij}$, and reflection
coefficients, $R_{ii}$. These coefficients are inserted in the
Landauer-B\"{u}ttiker formula in order to obtain the relevant resistances.
In the present work we will calculate those coefficients within
the semiclassical approach\cite{cBM, Ma1, cSCA2, cSCA3}. This
approach has been successfully used for mesoscopic semiconducting Hall bars with different potential and magnetic field profiles\cite{cSCA, cSCA4, cSCA5}. The problem is solved numerically by injecting a large number of particles (typically $10^5$) from each
lead of the structure. We consider carriers moving ballistically and
the billiard model \cite{cBM} is used to simulate their movement.

The paper is organized as follows. In Section \ref{Mod} our method of solving
the Hall bar problem is introduced and important symmetry relations for the transmission coefficients are pointed out. In Section \ref{Res} we present our numerical results for the Hall and bend resistances for a different values of the potential height and different positions of the pn-interface. A comparative study is made of the effect of shifting the pn-interface on the Hall and bend resistances and a comparison with the experiment of Ref. \onlinecite{Marcus1} is presented in Section \ref{sComp}.
Our conclusions and remarks are presented in Section \ref{Con}.
\section{Model}
\label{Mod}
To describe the transport properties of a graphene Hall bar the semiclassical billiard model\cite{cBM} is used. In this model electrons are considered as point particles (billiards) which are injected uniformly  over the length of the lead, while the angular distribution is given by $P(\alpha) = 1/2\cos(\alpha)$, with $\alpha \in [-\pi/2,\pi/2]$ . The model is justified when $\lambda_F\ll W$, where $\lambda_F$ is the Fermi wavelength and $W$ the width of the lead and when quantization effects are not important. It has been successfully used to describe various experiments containing a mesoscopic Hall bar\cite{A3,cBM,cSCA,cSCA5}.
The motion of ballistic particles is determined by the classical Newton equation of motion, which is justified for the case $l_\phi<W<l_e$ where $l_\phi$ is the phase coherence length and $l_e$ the mean free path (for a typical parameters at low temperatures the electron mean free path can be calculated as $l_e=(\hbar /e)\mu (\pi n_s)^{1/2}> 1\mu m$, with $\mu$ the mobility and $n_s$ the electron density), while the transmission of electrons through the potential step is calculated quantum mechanically using the Dirac  Hamiltonian. The trembling motion of electrons (Zitterbewegung) is neglected due to its transient character which makes it observable only on a femtosecond scale\cite{Zitt}. From the Dirac equation for graphene we obtain the relation $v_fp = mv_f^2= E_F-eV$, which we used in the classical equations of motion to substitute the mass term. In the presence of a magnetic field $B$ carriers move in circular orbits with cyclotron radius given by
\begin{equation}
r_c = \frac{|E_F-eV|}{ev_F|B|}
\end{equation}
where $V$ is the applied potential, $e$ the elementary charge, and $v_F$ the Fermi velocity.

The transmission of the particle through the potential step is treated quantum mechanically. Transmission and reflection probability are calculated according to the Dirac model. The transmission probability is given by
 \begin{equation}\label{Tf}
    T = \displaystyle{\frac{\cos{\alpha_i}\cos{\alpha_t}}{ \cos^{2}{((s_{1}\alpha_i + s_{2}\alpha_t)/2)}}},
\end{equation}
where $\alpha_i$ and $\alpha_t$ are incident and transmitted angle, respectively, $s_1 = sgn(E_F-eV_i)$ and $s_2 = sgn(E_F-eV_t)$. In our simulation this is treated in the following way: each time a particle approaches the potential step the transmission probability is compared with a random number generated from a uniform distribution. If the transmission probability is smaller than a generated number the particle will transmit, otherwise it will be reflected. In the later case the reflected angle is the same as the
incident one. In case of transmission incident and transmitted angle
are connected with Snell's law:
\begin{equation}\label{eq1}
\sin\alpha_t = \left(\frac{E_F-eV_i}{E_F-eV_t}\right)\sin\alpha_i,
\end{equation}
where $V_i$ is the potential of the region from which the particle is transmitted to a new region with
potential $V_t$.

We are interested to examine the behavior of the Hall resistance $R_H = R_\alpha = V_{\alpha}/I_{\alpha}$ and the bend resistances $R_B = R_\beta= V_{\beta}/I_{\beta} $, for electron current and voltage measured as in Fig. \ref{fig1}. The position of the pn-interface with respect to the Hall bar is taken arbitrary. The resistances are calculated according to the Landauer-B\"{u}ttiker theory where the resistances for a four terminal cross shaped structure are defined as
\begin{equation}
R_{mn,kl} = R_0\frac{T_{km}T_{ln}-T_{kn}T_{lm}}{D},
\end{equation}
where D is a subdeterminant  of the transmission matrix $T$ which is symmetric in the magnetic flux:\cite{cBUT} $D(B) = D(-B)$ given by
\begin{equation}
D = \alpha_{11}\alpha_{22}-\alpha_{12}\alpha_{21},
\end{equation}
where
\begin{equation}
\begin{array}{l}
\displaystyle{\alpha_{11} = \left[(T_{21}+T_{31}+T_{41})S-(T_{14}+T_{12})(T_{41}+T_{21})\right]/S} \\
\displaystyle{\alpha_{12} = (T_{12}T_{34}-T_{14}T_{32})/S} \\
\displaystyle{\alpha_{21} = (T_{21}T_{43}-T_{41}T_{23})/S }\\
\displaystyle{\alpha_{22} = \left[(T_{12}+T_{32}+T_{42})S - (T_{21}+T_{23})(T_{32}+T_{21})\right]/S}\\
\displaystyle{S = T_{12}+T_{14}+T_{32}+T_{34}.}\\
\end{array}
 \end{equation}

Transmission coefficients $T_{ij}$ are the probability that the electron
injected from lead $j$ will end up in lead $i$. The presence of the pn-interface breaks
the four-fold symmetry of the system and we find that the Onsager relation are no longer valid but some symmetry relations still survive:  $T_{j2}(B) = T_{j4}(-B)$ and $T_{2j}(B) = T_{4j}(-B)$, with $j = \lbrace 1,3 \rbrace $, $T_{24}(B) = T_{42}(-B)$, $T_{13}(B) = T_{13}(-B)$ and $T_{31}(B) = T_{31}(-B)$. The two last one are important for the bend resistances. Using them we derive
\begin{equation}
\begin{array}{l}
\displaystyle{R_{14,32}(B) \propto T_{31}(B)T_{24}(B)-T_{34}(B)T_{21}(B) }=\\
\displaystyle{ T_{31}(-B)T_{42}(-B)-T_{32}(-B)T_{41}(-B) \propto -R_{12,43}(-B).}
\end{array}
\end{equation}
In a similar way we obtain $R_{32,14}(B) = -R_{43,12}(-B)$.
All bend resistances can be connected if one approximation is made which are justified by our numerical simulations. Transmissions $T_{31}$ and $T_{13}$ give us the percentage of carriers that are able to transmit through the potential step and end up in the opposite lead. Due to the presence of the pn-interface we have $T_{31}(B)\neq T_{13}(-B)$. But it is obvious that these probabilities are nonzero only for small values of the magnetic field, otherwise the cyclotron radius of carriers is too small to overcome the width of the lead. At small fields carriers move practically along straight lines and due to the Klein tunneling in graphene the highest transmission through the potential step will be for carriers that approach the pn-interface perpendicularly. For these carriers transmission to the other region will not change their path in a significant way and we have approximately $T_{31}(B)\approx T_{13}(-B)$. For a perfect symmetric Hall bar we can write $T_{34}(B)T_{21}(B) \approx T_{43}(-B)T_{12}(-B)$ which we use to derive $R_{14,32}(B) \approx R_{32,14}(-B)$. Finally, we can write $R_{14,32}(B) \approx R_{32,14}(-B)\approx -R_{12,43}(-B)\approx -R_{43,12}(B) $.

Using the fact that for a Hall bar with symmetric leads  $T_{23}(B)T_{41}(B) = T_{32}(-B)T_{14}(-B)$ holds, we have $R_{13,24}(B) = R_{24,13}(-B)$.
Therefore, it is sufficient to calculate only one Hall resistance $R_H = R_{13,24}$ and one bend resistance $R_B = R_{14,32}$.

Following dimensionless units are used: $B^* = B/B_0$, with $B_0 =|E_F|/ev_FW$ and $R^* = R/R_0$ with $R_0 = (h/4e^2)(\hbar v_F/|E_F|W)$, where $W$ is the width of the Hall bar, $E_F$ is the Fermi energy, $e$ electron charge, $v_F$ is Fermi velocity.
For our numerical calculations we used the following typical parameters, $W = 1 \mu m$, $E_F = 50 meV$, $v_F = 10^6 m/s$, and a time step $\Delta t = 0.005W/v_F$. This results into $R_0 = 0.0132 \frac{h}{4e^2}$ and $B_0 = 0.05$ T.
\section{Results and discussion}
\label{Res}
In the following discussion we will assume that $V_1 = 0$ and only the
potential of the second region is changed, i.e. $V_2 = V$. For simplicity we assume a perfect symmetric Hall bar with $W=L$.

\begin{figure}[h]
\begin{center}
\includegraphics[width=6cm]{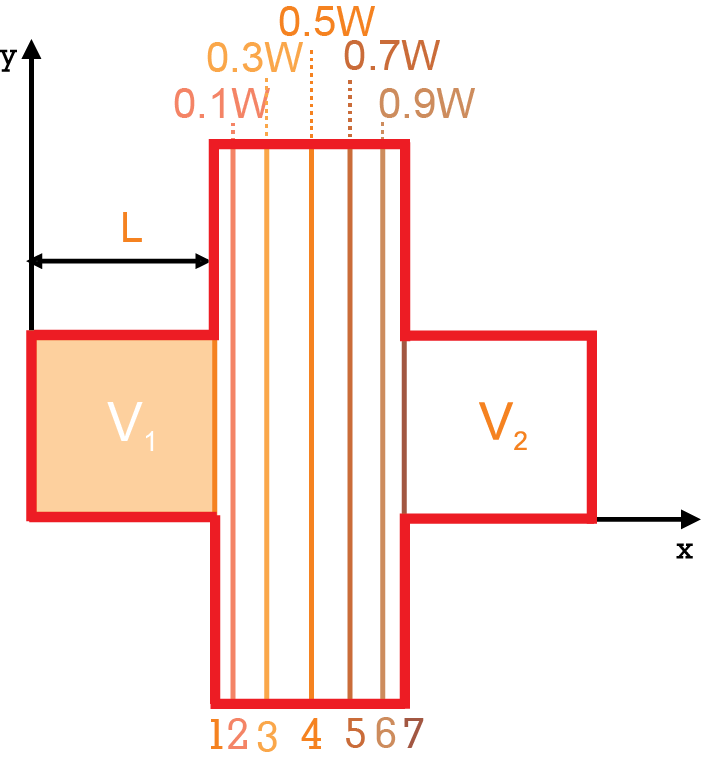}
\caption{(Color online) Schematic of the Hall bar with the pn-junction placed at $D = L$ and indication of the different positions of the pn-interface.}
\label{fig10}
\end{center}
\end{figure}

\begin{figure}[htbp]
\begin{center}
\includegraphics[width=8.5cm]{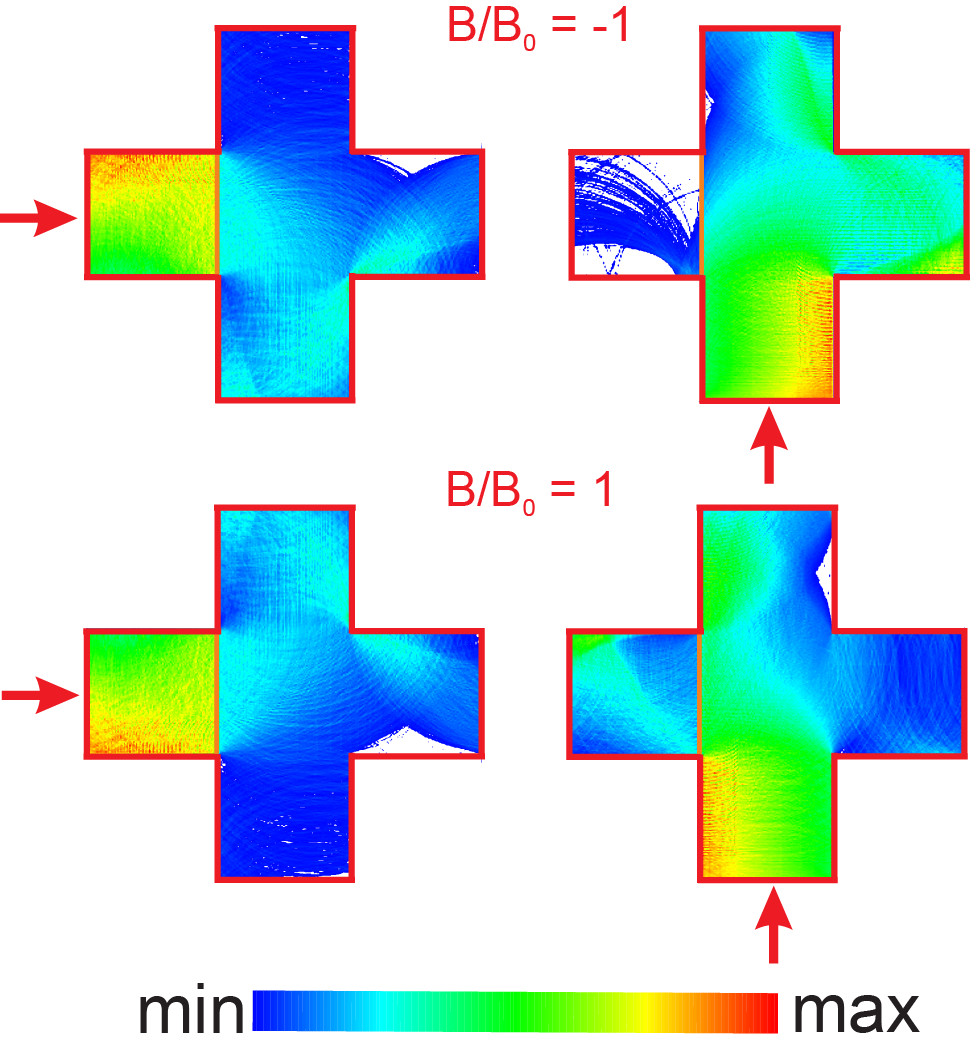}
\caption{(Color online) Electron densities for the pn-interface placed at $D = L$ and $eV/E_F = 2$. Values of applied magnetic field are shown in the figure. Arrows indicate the lead of injection.}
\label{fig_ed_10}
\end{center}
\end{figure}

First, we study the case when the pn-junction is placed at the end of terminal $1$, at $D = L$, as shown in Fig. \ref{fig10}.
We see that potential $V_1=0$ is applied only on lead $1$, while the rest of the Hall bar is at potential $V_2=V$. To have a complete grasp of the electron motion in this system, electron current density plots of the injected electrons are used. These plots show us the flow density of the current carriers in the structure and help us to better understand the behavior of the resistances as function of magnetic field. Fig. \ref{fig_ed_10} shows electron densities for $eV/E_F = 2$ and two different values of the magnetic field and injection of electrons through two different leads.
This value of applied potential is chosen because it leads to the same cyclotron radius of carriers in  both regions of the structure.
\begin{figure}[htbp]
\begin{center}
\includegraphics[width=6.5cm]{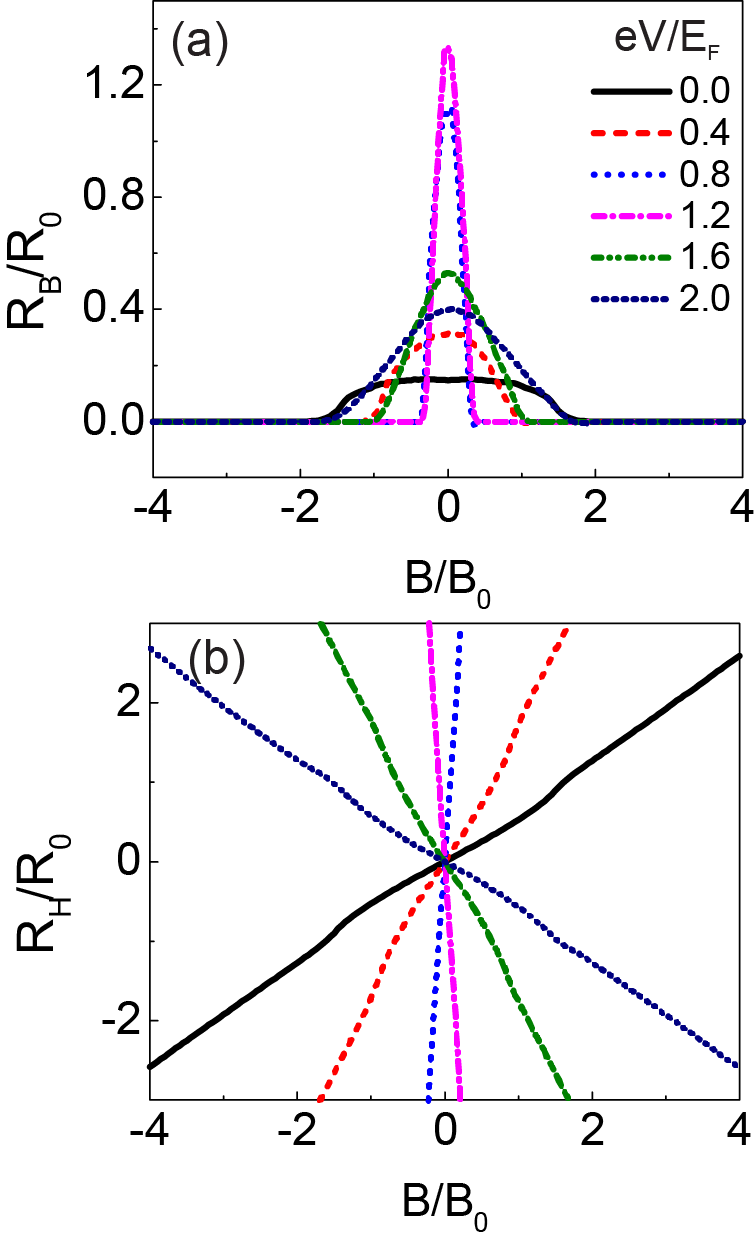}
\caption{(Color online) Plots of (a) the bend and (b) the Hall resistance in case when the pn-interface is at $D = L$. Plots are made for different values of the applied potential in the second region (or equivalently different relative position of the Fermi energy). Values of $V$ scaled with Fermi energy are shown in the inset of (a).}
\label{fig112}
\end{center}
\end{figure}
From Fig. \ref{fig_ed_10} we see that the sign reversal of the magnetic field is not affecting the transmission probabilities $T_{31}$ and $T_{24}$. For small magnetic field most of the electrons will have cyclotron orbits larger than the width of the lead and therefore will be able to reach the opposite lead. Changing the sign of the applied field for which $r_c>W $ will not affect transmissions between opposite leads in a significant way. On the other hand in this configuration leads 2, 3 and 4 are subject to the same potential. This is important because now we will have that $T_{34}(B)T_{21}(B)\approx T_{34}(-B)T_{21}(-B)$. Due to this the bend resistance will be symmetric with respect to the sign reversal of the applied magnetic field.
Another important point for this system is that although we have a pn-junction snake states don't appear in this structure. Reason for this is the fact that the pn-interface is present only in the first lead and not through the whole length of the structure. Plots of resistances are shown in Fig. \ref{fig112}. First observation is that the bend resistance is indeed symmetric with respect to a sign change of the magnetic field. Plot of $R_B$ shows that the peak around $B = 0$ is higher as we approach zero kinetic energy in region $2$. The same thing we can say for the plot of the Hall resistance, $R_H$, the slope of the curves is steeper for smaller values of the kinetic energy in region 2. Reason for this kind of behavior lies in the fact that transmission is higher for smaller incident angles (perfect transmission for normal incidence - Klein tunneling) and therefore carriers that do transmit will most likely end up in lead 3 which enhance probability $T_{31}$ and at the same time the bend resistance, $R_B$. For higher magnetic fields carriers make smaller circles which results into larger incident angles on the pn-interface. The transmitted angle, $\alpha_t$, given by Eq. \eqref{eq1} is larger for smaller vales of $(E_F-eV)$ and therefore for higher fields the majority of carriers will be reflected back or scattered to the perpendicular leads. This is why the range of applied magnetic field for which $R_B$ is different from zero becomes smaller as $(E_F-eV)$ approaches zero.
A shoulder in the plots of the Hall resistance can be spotted at magnetic fields for which the bend resistance changes from nonzero to zero value and vice versa. These are the values for which the cyclotron radius is equal to the width of the lead emphasizing the fact that carriers are no longer able to reach the opposite lead and the injected current will only flow towards the perpendicular leads.

\begin{figure}[htbp]
\begin{center}
\includegraphics[width=6.5cm]{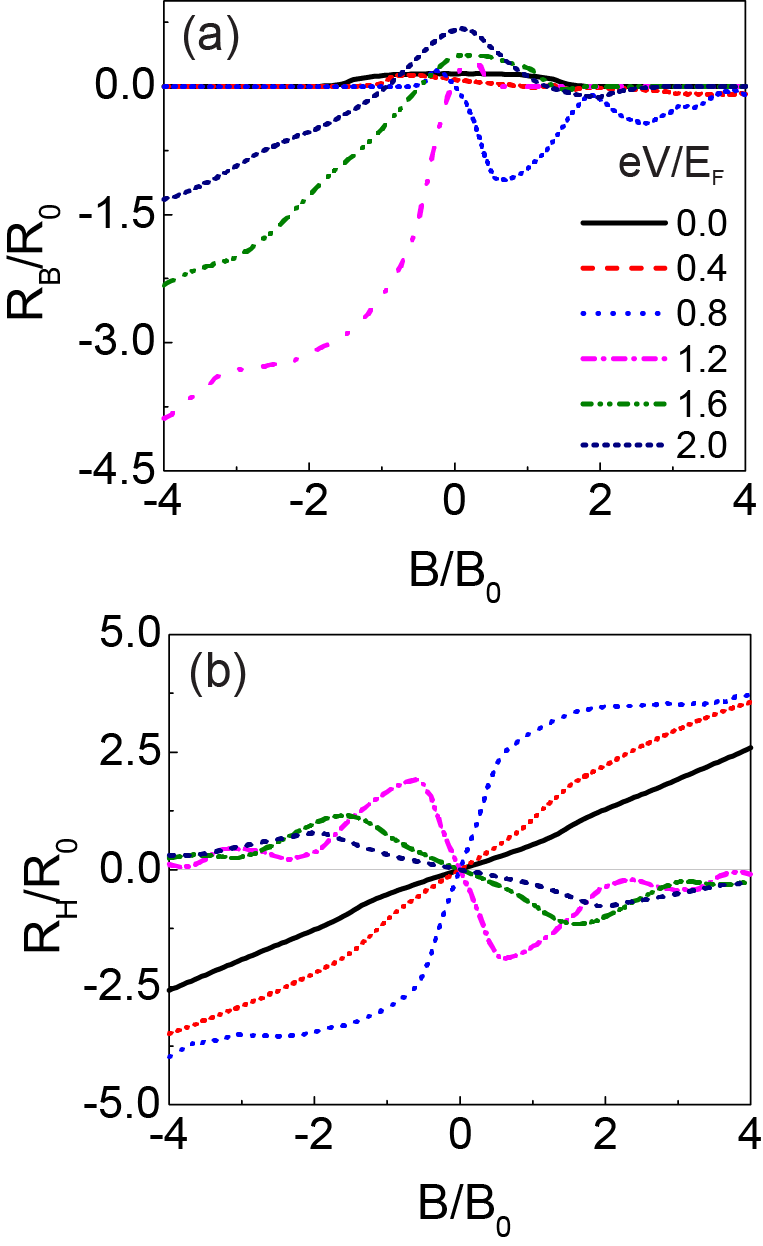}
\caption{(Color online) The same as Fig. \ref{fig112} but now for a pn-interface at $D = L+0.3W$.}
\label{fig17}
\end{center}
\end{figure}
Next, let's look at the case when $D = L+0.3W$, as shown in Fig. \ref{fig10} at position 3.  Now, there is a gap between the left edge of lead $2$ and the pn-interface which means that there is the possibility for the occurrence of snake states. Plots of resistances are shown in Fig. \ref{fig17} which are very different from the previous case of Fig. \ref{fig112}. We notice that the bend resistance is increased, while the Hall resistance is smaller. These plots resemble the ones obtained for the case $D = L+0.5W$ (given in Ref. \onlinecite{A3}), when the pn-interface is in the middle of the Hall bar. Asymmetric behavior of the bend resistance for $eV/E_F>1$ can be spotted while the Hall resistance exhibits saturation for $eV/E_F<1$. Compared with the case in Ref. \onlinecite{A3} the bend resistance is smaller for zero magnetic field. The peak around $B=0$ is a consequence of the Klein tunneling \cite{f3, f5} and the focusing effect\cite{f4} in graphene. For the pn-interface further away from lead 3 the focusing effect is less expressed and the electrons are widespread in the second region which has an important influence on the bend resistance. We see that the Hall resistance in the pn-regime is nonzero for a broad range of magnetic field values but it is not hard to conclude that for large fields it will go to zero because the cyclotron radius is small enough that the currents from lead 1 and 3 will end up in lead 4 for positive $B$ or in lead  2 for negative values of $B$ ($R_H \propto T_{23}T_{41}-T_{21}T_{43}$), while in the nn-regime the results will coincide with those for case $V = 0$.
\begin{figure}[htbp]
\begin{center}
\includegraphics[width=8.5cm]{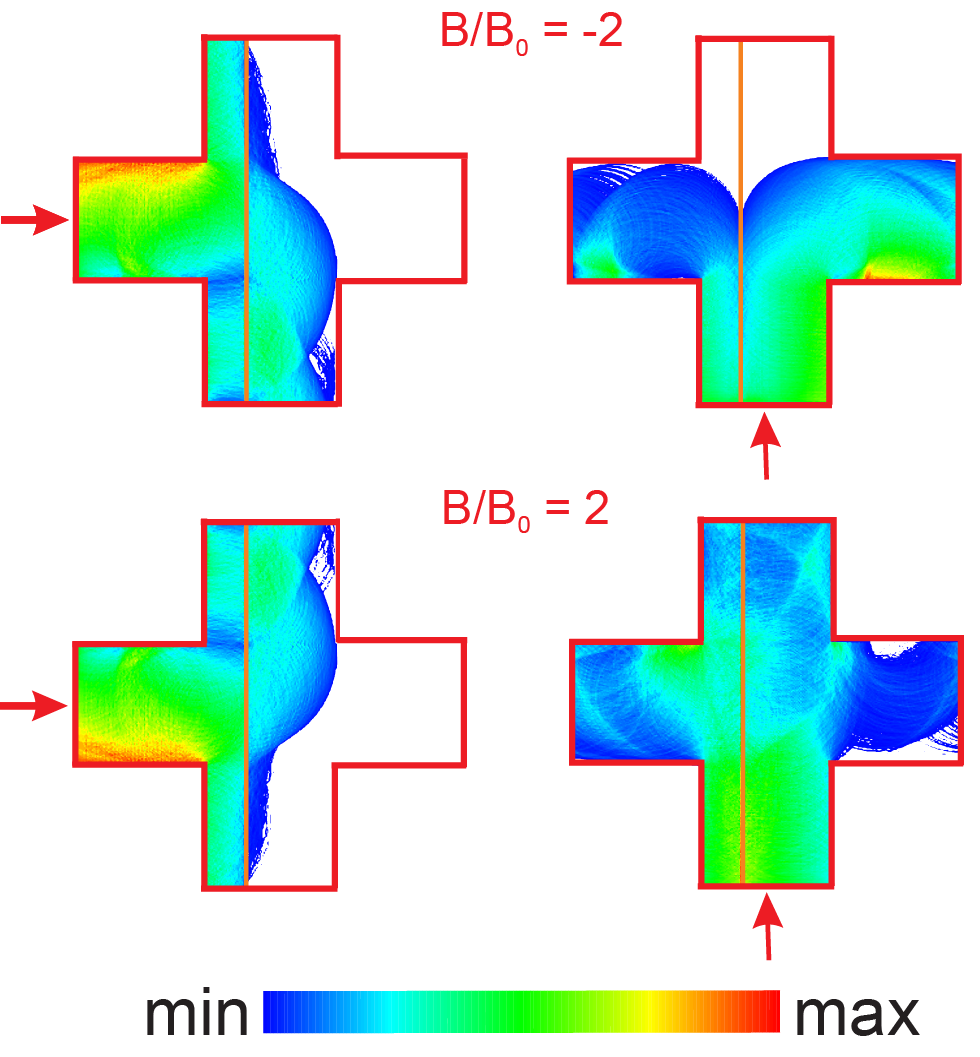}
\caption{(Color online) Electron density for pn-interface placed at $D = L+0.3W$ and $eV/E_F = 1.2$. Values of applied magnetic field are shown in the figure. Arrows indicate the lead of injection.}
\label{fig16}
\end{center}
\end{figure}
To better understand the behavior of the bend resistance we will again use electron density plots.
Plots of electron current flow in this system for two values of magnetic field and $eV/E_F = 1.2$ are shown in Fig. \ref{fig16}. Injection from lead 4 in case of $B/B_0 = 2$ shows the presence of snake states in this configuration. Snake orbits should enhance the bend resistance but as we see for positive values of $B$ the bend resistance is very small. From the relation $R_B\propto T_{31}T_{24}-T_{34}T_{21}$ this is easily understandable. Although $T_{24}$ is very high $T_{31}$ for this case is zero which results in a small negative value for $R_B$. For negative magnetic field we have a different situation. Now $T_{24}$ is zero but the majority of electrons is bent to the perpendicular leads which results in high negative values for $R_B$.

\begin{figure}[htbp]
\begin{center}
\includegraphics[width=6.5cm]{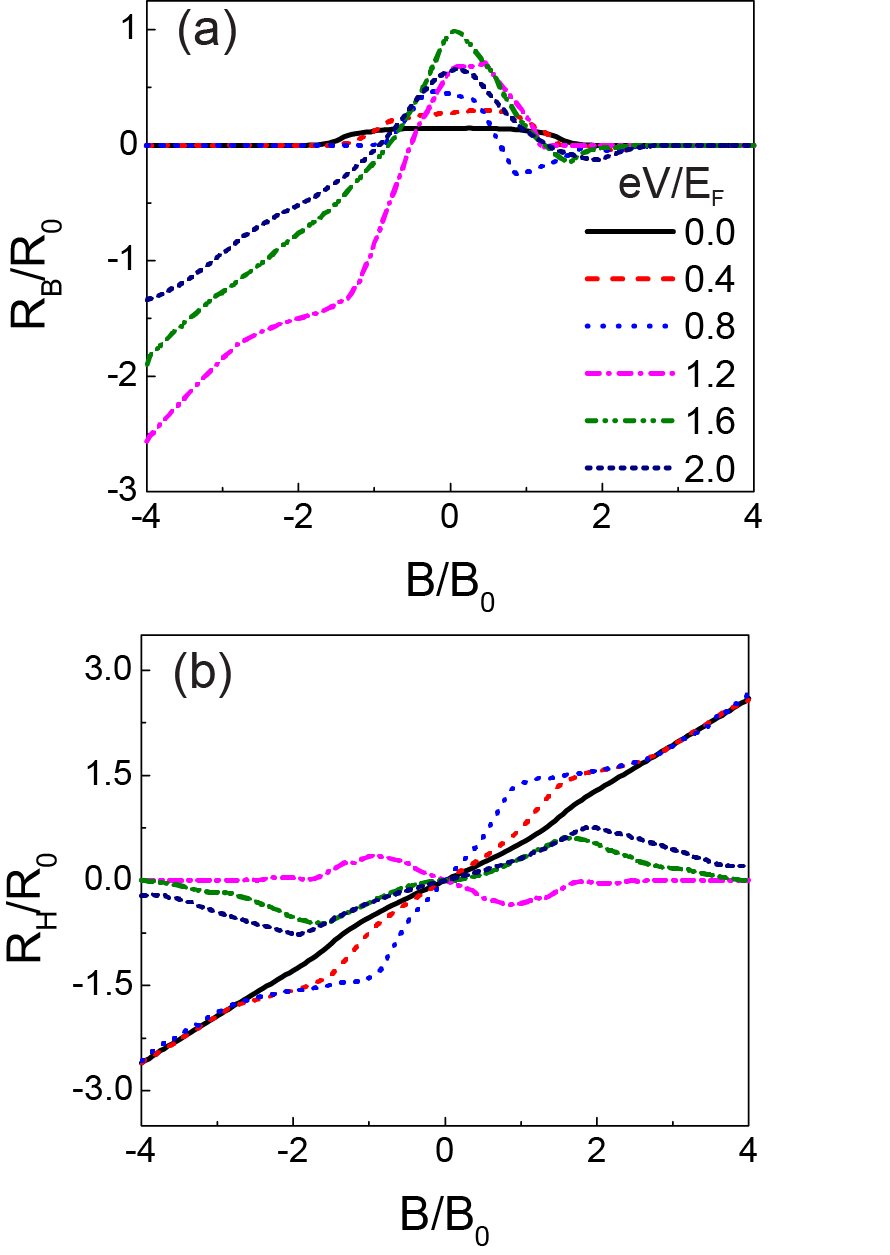}
\caption{(Color online) The same as Fig. \ref{fig112} but now for a pn-interface at $D = L+0.7W$.}
\label{fig21}
\end{center}
\end{figure}

\begin{figure}[htbp]
\begin{center}
\includegraphics[width=8.5cm]{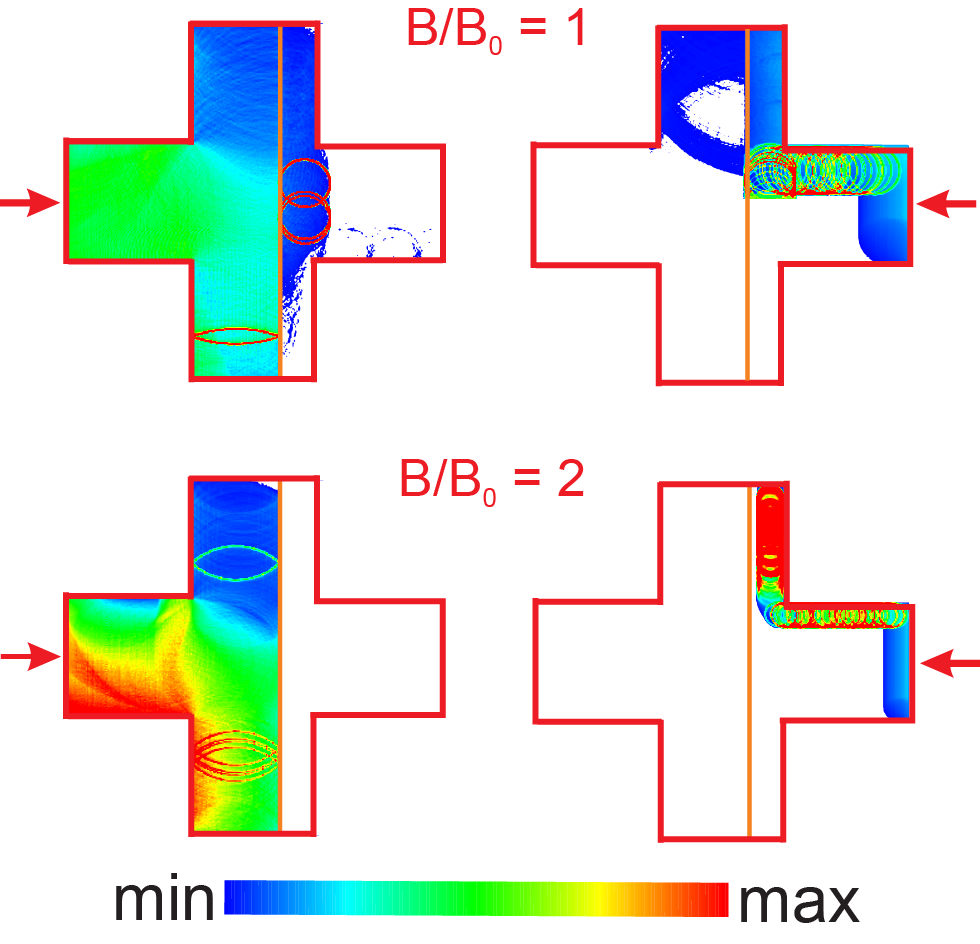}
\caption{(Color online) Electron density for the pn-interface placed at $D = L+0.7W$ and $eV/E_F = 0.8$. Values of applied magnetic field are shown in the figure. Arrows indicate the lead of injection.}
\label{fig20}
\end{center}
\end{figure}

Now, we will shift the pn-interface to the right of the central position, at the value $D = L+0.7W$, at shown in Fig. \ref{fig10} as position 5. The area under potential $V_2$ is now smaller than the one under potential $V_1$.
Plots of resistances are shown in  Fig. \ref{fig21}. The peak around zero magnetic field is more pronounced as compared to the previous case. This can be explained with the fact that the electron beam in the second region is well focused, directing most of the electrons to lead 3. The bend and the Hall resistance go faster to zero when a positive magnetic field is applied which again can be explained by the fact that the pn-interface is further away from lead 1 which means that electrons injected from this lead are unable to cross it when a higher magnetic field is applied.
To understand the occurrence of the Hall plateau-like features we show electron current density plots for two different values of the magnetic field in Fig. \ref{fig20}. First, we observe the injection from lead 1. In the case $B/B_0 = 1$ there is a relatively small percentage of electrons that transmit through the pn-interface. This is because the pn-interface is far away for electrons to reach it with small angles and Eq. \eqref{eq1} tells us that only the electrons that have $\alpha_i<\arcsin(1/5)$ can pass through, therefore most of them will be reflected back. If we increase the magnetic field to $B/B_0 = 2$ there are no electrons that can transmit to the opposite terminal. Injection from  terminal 3 shows a different behavior. Now, the cyclotron radius in this region is very small and all the electrons will end up in the same lead. So, we see that in this range of magnetic field the change of its value will not affect the Hall resistance in a significant manner and hence a plateau-like behavior develops.

\begin{figure}[htbp]
\begin{center}
\includegraphics[width=6.5cm]{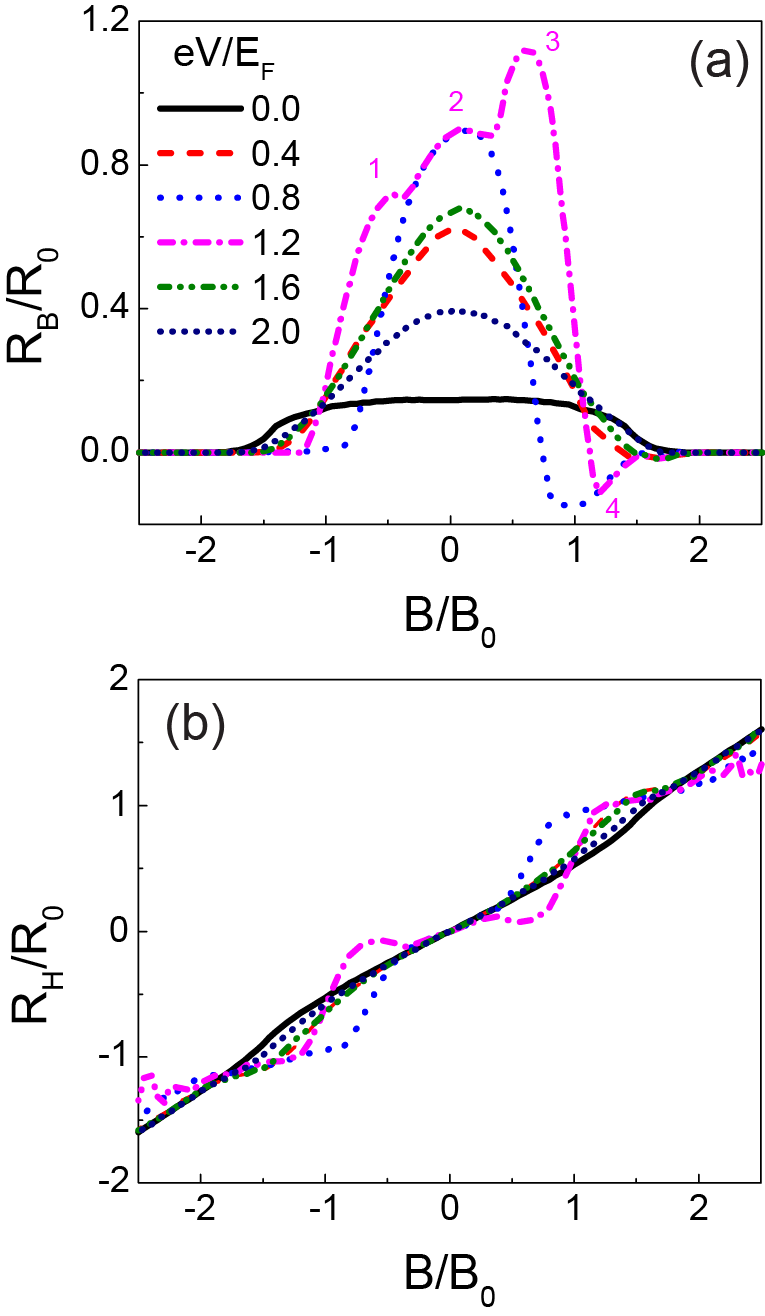}
\caption{(Color online) The same as Fig. \ref{fig112} but now for a pn-interface at $D = L+W$.}
\label{fig31}
\end{center}
\end{figure}

\begin{figure*}[!]
\begin{center}
\includegraphics[width=11cm]{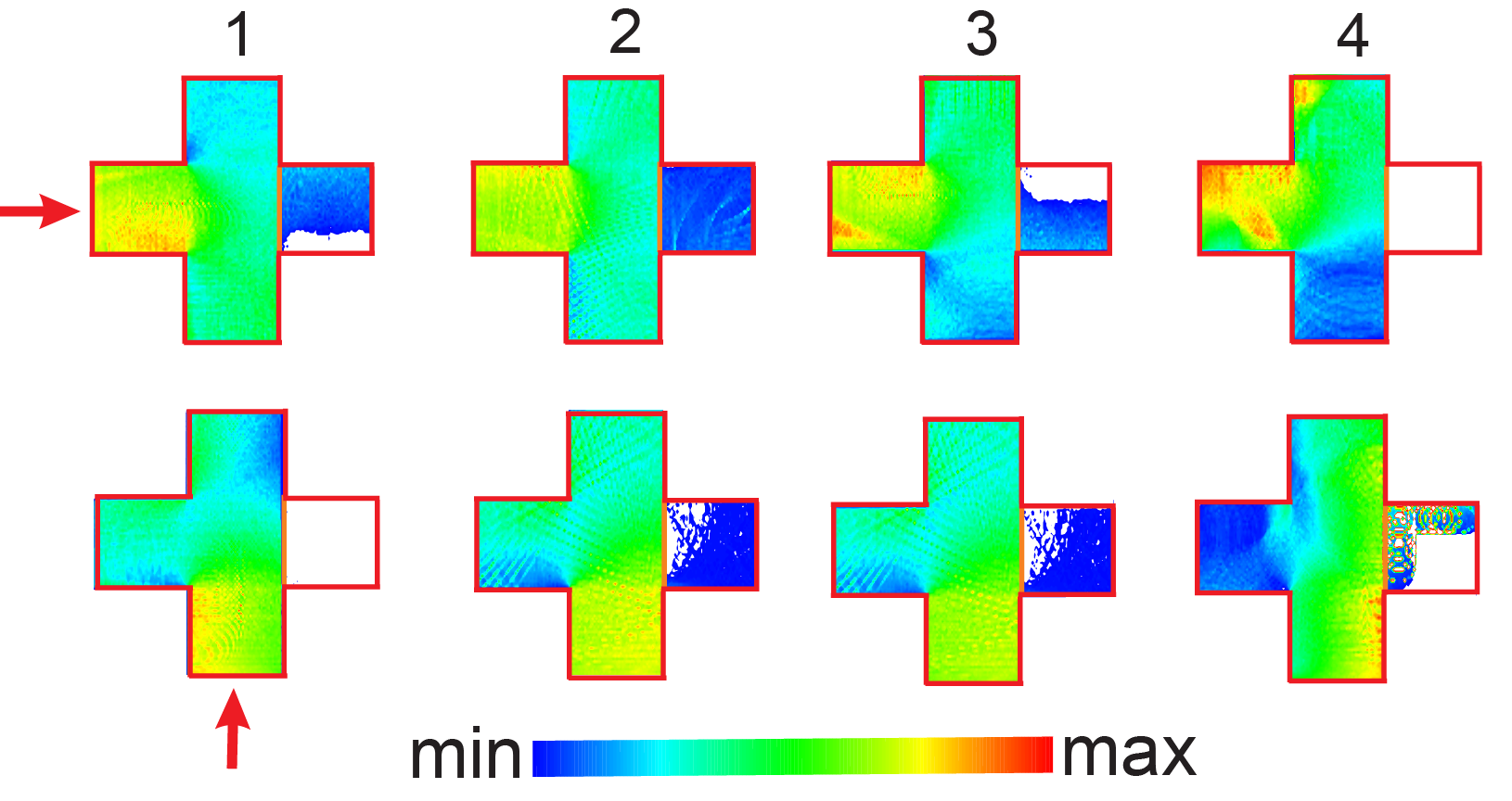}
\caption{(Color online) Electron density for points indicated in Fig. \ref{fig31}(a) for the pn-interface placed at $D = L + W$ and $eV/E_F = 1.2$. The first row shows injection from lead 1 and the second row shows injection from lead 4.}
\label{fig28}
\end{center}
\end{figure*}

Finally, the pn-interface is set to the beginning of lead $3$. This situation is shown in Fig. \ref{fig10}, position 7. Now, the potential $V$ is only present in lead $3$ while the rest of the Hall bar is under potential $V_1 = 0$. Plots of resistances are given in Fig. \ref{fig31}.  Interesting point is that asymmetric behavior of the bend resistance isn't present in this plot which means that there are no snake states occurring in this structure. We notice the change of sign of the bend resistance for values of magnetic field for which a Hall plateau-like structure occurs which happens as a consequence of an incomplete guiding of carriers\cite{cSCA}. Another difference is that the Hall resistance shows similar behavior for all values of applied potential $V_2=V$.
In this configuration we see that for $eV/E_F = 1.2$ the bend resistance has an unusual shape. Maximum value occurs for nonzero magnetic field. To understand this kind of behavior we show electron current density plots in Fig. \ref{fig28} for four points marked in Fig. \ref{fig31} on the dash-dotted curve. For point 1 we see that $T_{34}$ is zero, which means that $R_B\propto T_{31}T_{24}$. Magnetic field bends the electrons in such a way that they are unable to penetrate through the potential step. If we increase the magnetic field we come to point 2. Now, we are around zero magnetic field and the electron paths are almost straight lines which means that transmission toward the opposite lead will increase and hence the bend resistance also. If we increase the magnetic field even further we are at point 3. From Fig. \ref{fig28} we see that $T_{24}$ is almost the same as for points 2 and 3, but the change comes from $T_{31}$. For $eV/E_F = 1.2$ and $B/B_0\rightarrow 1$ the cyclotron radius is approaching $W$. This means that the highest percentage of electrons injected from lead 1 will reflect once on the border of terminal 1 and arrive to the pn-interface with relatively small angle and will pass through. Transmission $T_{31}$ reaches its maximum for this value of magnetic field. Therefore, the bend resistance will increase further. If we continue increasing the magnetic field we reach $r_c = W$ but in this case we have a negative bend resistance which occurs as a consequence of high $T_{34}$ and $T_{31} = 0$. Cyclotron radius is now equal to the width of lead $W$ and electrons injected from lead 4 will have a higher probability to be scattered to lead 3, while the transmission of electrons injected from lead 1 is minimum.
\section{Comparison}
\label{sComp}
\begin{figure*}[htbp]
\begin{center}
\includegraphics[width=12cm]{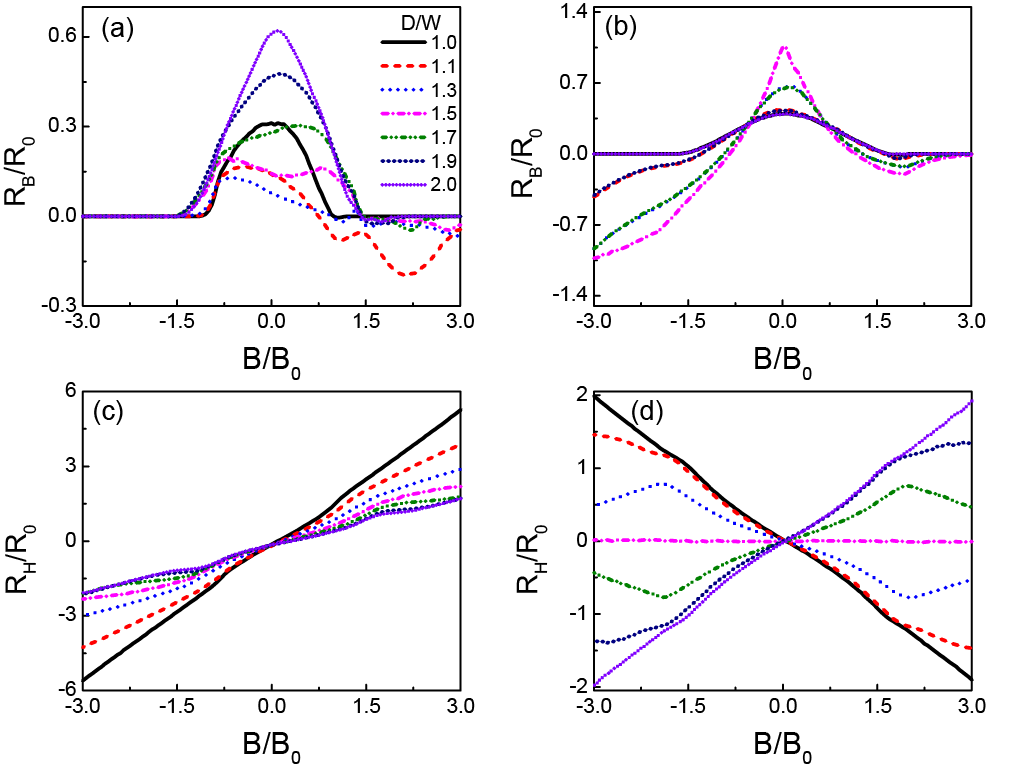}
\caption{(Color online) (a) The bend and (c) the Hall resistance versus magnetic field for applied potential $eV/E_F = 0.4$. (b) The bend and (d) the Hall resistance versus magnetic field for applied potential $eV/E_F = 2$.}
\label{fig29}
\end{center}
\end{figure*}

Results for a symmetric positioned pn-junction were obtained earlier in Ref. \onlinecite{A3}. We generalized the results of Ref. \onlinecite{A3} to an asymmetric location of the pn-junction within the Hall bar.
We will compare now the bend resistance, $R_B$, and Hall resistance, $R_H$, for different positions of the pn-junction interface. Plots show how the resistances are changing as we shift the pn-interface from the end of terminal $1$ to the beginning of terminal $3$. These plots are shown in Fig. \ref{fig29}. Values of shifts are the same in all plots and for $W=L$ these are: $W$, $1.1W$, $1.3W$, $1.5W$, $1.7W$, $1.9W$ and $2W$.
\begin{figure}[htbp]
\begin{center}
\includegraphics[width=6.5cm]{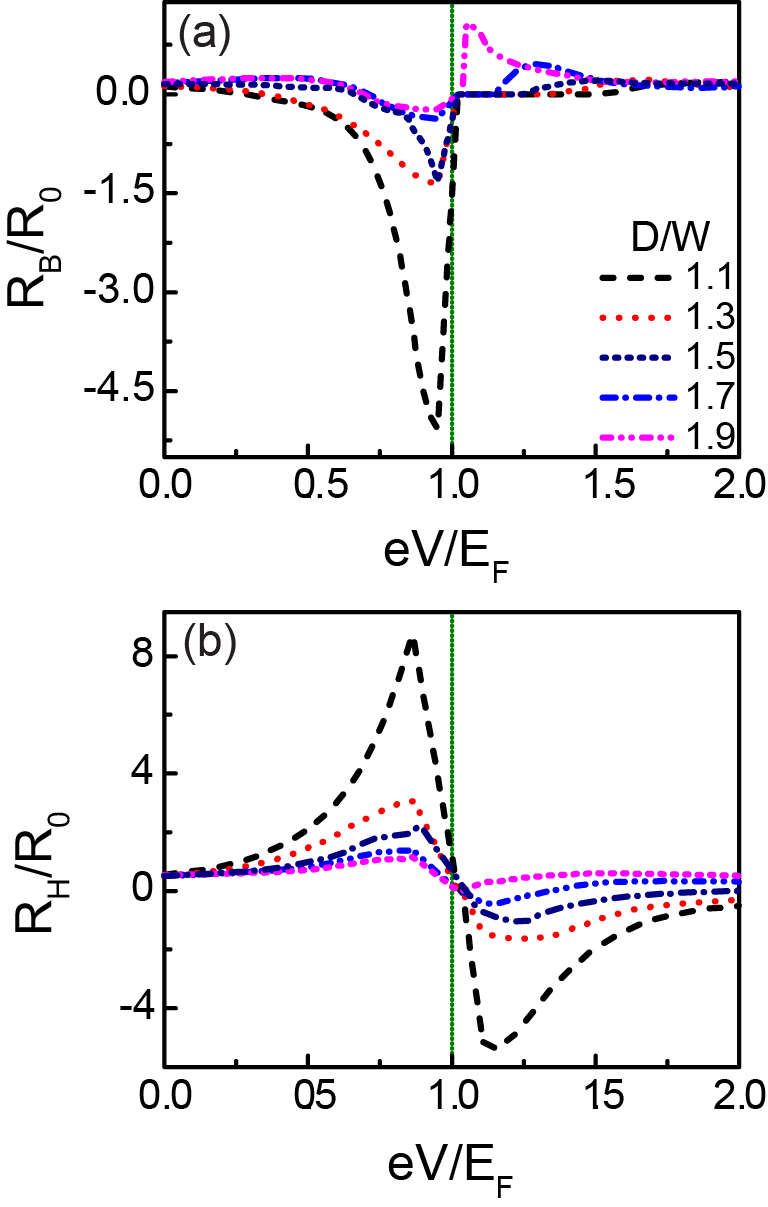}
\caption{(Color online) (a)The bend and (b) the Hall resistance versus the potential $eV/E_F$ for different positions of the pn-interface. Plots are made for $B/B_0 = 1$ while the green vertical line shows the transition between nn- and np-regime.}
\label{comparison}
\end{center}
\end{figure}
Fig. \ref{fig29} (a) and (b) show qualitative different behavior of the bend resistances when the second region is n- or p-type. In the first case we see that as we go from the end of terminal 1 to the beginning of terminal 3 values of $B$ for which the resistance is different from zero decreases, while the peak around $B = 0$ increases. This happens because the electrons that transmit through the potential step will have angles that are larger than the incident angle($|E_F|>|E_F-V|$) and for the pn-interface far away from terminal 3 they will be scattered to leads 2 or 4. If we put the pn-interface closer to terminal 3 the number of electrons that end up in this terminal is higher, which increases $R_B$ ($R_B\propto T_{31}T_{24}-T_{34}T_{21}$). This behavior of resistances excludes case $D = L$ where now the pn-interface exists only in the middle part of the Hall bar and not as usual in leads 2, 4 and middle part. Having this in mind we know that for small $B$ there are electrons injected from lead 4 which won't be scattered by the pn-junction at all, allowing them to end up in terminal 2. This increases the resistance significantly. Notice the minimum in the bend resistance for nonzero magnetic fields, their absolute value decrease as the pn-interface is shifted away from terminal 1. This happens because the cyclotron radius in the second region is much smaller than in region 1, which means that the electrons from the second region can't reach the pn-interface, while in the other region electrons that pass the potential step will decrease its cyclotron radius and won't be able to reach terminal 3. As a consequence $T_{31}T_{24}$ is very small, while $T_{34}T_{21}$ increases as the pn-interface is shifted away from terminal 3 resulting in high and negative $R_B$. The second case shows the asymmetric behavior of $R_B$ for magnetic fields of opposite sign.

Hall resistance is shown in Figs. \ref{fig29} (c) and (d). Again, all the features are preserved as we shift the pn-junction interface through the Hall bar. For $eV/E_F<1$ we have plateau-like behavior in the resistance plots. If the interface is placed around terminal 1 the plateau-like feature is almost invisible and as we approach terminal 3 it is more noticeable. The plateau occurs because of the pn-interface and the fact that the transmission through the potential step is restricted to a range of allowed angles, which doesn't change with $B$. Plateaus are more distinct for energies around zero because for those values the range of allowed angles is very small and therefore this effect is more expressed. For $eV/E_F>1$ the behavior of the resistance is different. For zero magnetic field $R_H$ is zero and then it starts increasing up to some value after which it decreases to zero. From the fact that $R_H\propto T_{23}T_{41}-T_{21}T_{43}$ this is easily understood. Because $E_F<eV$ one side of the bar is n-type and the other p-type which means that electrons that are on different sides of the potential step will move in opposite directions. This tell us that if on one side $T_{41}$ increases on the other side $T_{43}$ will do the same. The first one makes $R_H$ larger while the other one lowers it. For a perfectly symmetric situation $T_{41} = T_{43}$ the Hall resistance is always zero as is the case for $eV/E_F = 2$ and when the pn-junction interface is placed in the middle. For the other positions of the pn-interface, depending on the size of the cyclotron radius in both regions, the position of the peak is shifted.

 Figs. \ref{fig29}(b) and \ref{fig29}(d) show the resistances for $eV/E_F = 2$.  For this case the cyclotron radius is the same in both regions. Bend resistances are equal for equal shifts to the left and to the right from the central position, while the Hall resistances have the same absolute value but its sign for opposite shifts are different.

Finally, we compare the bend and the Hall resistance with the experimental results from Ref. \onlinecite{Marcus1}. Behavior of $R_H$ is investigated for different values of the potential $V_2=V$ when the pn-interface is placed in the middle of the Hall bar in the presence of a magnetic field $B/B_0 = 1$. $V$ is chosen in such a way that a transition from the nn-regime to the np-regime occurs and the charge-neutrality point (CNP) is shown with a green vertical line in Fig. \ref{comparison}. We see that when we cross the CNP, both $R_H$ and $R_B$ change sign. Behavior of the Hall resistance shows a good agreement with the experimental results presented in Fig. 2(c) of Ref. \onlinecite{Marcus1}. The bend resistance exhibits a peak in the nn-regime for values of $eV$ around $E_F$. The value of the peak is exceptionally high for the case when the pn-interface is positioned at $D=1.1W$ and decreasing as we move closer to lead 3. This peak can be explained using the fact that transmission through the potential step is decreasing significantly as $E_F-eV\rightarrow 0 $ together with the very small cyclotron radii for electrons that do transmit. This has a major impact for the injection from lead 1. If the pn-interface is further away from lead 3 electrons are practically unable to reach lead 3 and  $T_{31}\rightarrow 0$, which leads to $R_B\propto -T_{34}T_{21}$. Of course, as we move the interface closer to terminal 3 or lower the height of the potential step $T_{31}$ will be different from zero and
the bend resistance will regain the former dependence. For small values of $E_F-eV$ we
can say that $T_{43}$ will go to zero for the same reasons as $T_{31}$. This will leave $R_H \propto T_{23}T_{41}$. Unlike $T_{34}$ and $T_{21}$, transmission $T_{43}$ will not be the highest for $E_F-eV\rightarrow 0$ because this will mean that most of the electrons injected from lead 3 will scatter back to this lead, but for some value different from 0 but still small enough that electrons are unable to reach the pn-interface. This is why peaks for $R_H$ are shifted away from CNP.
\section{Conclusion}
\label{Con}
We have studied the electronic response of a four terminal graphene Hall bar with an asymmetric
pn-junction. Properties of the structure were examined using the bend resistance and the Hall resistance. Two different regimes are observed: $i)$ when both regions are n-type, and $ii)$ when
the second region is p-type. The first case shows that the Hall resistance dominates and Hall plateau-like behavior is found. For the nn-interface positioned further away from lead $3$ the plateau feature is hardly visible but becomes more noticeable as we approach lead $3$. This was explained with the
fact that transmission through the potential
step is restricted by Snell's law on a finite range of angles and the Hall bar will guide electrons to the perpendicular leads. In the pn-regime the bend resistance dominates. The asymmetric behavior of it occurs due to an additional conduction mechanism - snake states whose movement are uni-directional and which have a significant influence on the resistance. Therefore, measuring the magnetic field dependence and size of $R_B$ should be a direct way to detect the presence and the effect of the snake states on the conduction. For the pn-interface placed at the end of terminal $1$ or the beginning of lead 3 snake states don't appear and $R_B = 0$. Moving the pn-interface in the Hall bar the peak around $B=0$ in the plot of the bend resistance is also changing. The reason for this was found in the fact that the focusing effect that appears due to the metamaterial properties of graphene in this regime is influenced greatly by the position of the pn-interface. Finally, we investigate the behavior of the Hall and the bend resistance for different  values of applied potential $V_2=V$. The bend resistance exhibits a peak around the charge-neutrality point which is not changing its position when the pn-interface is displaced. The Hall resistance on the other hand changes its sign when the CNP is crossed and has a maximum for a potential slightly shifted away from the CNP. This is in agreement with he recent experiment of Ref. \onlinecite{Marcus1}.

\section{Acknowledgment}
This work was supported by the Flemish Science Foundation (FWO-Vl), the European Science Foundation (ESF) under the EUROCORES
Program EuroGRAPHENE within the project CONGRAN and the Methusalem Foundation of the Flemish government. We acknowledge fruitful discussions with M. Barbier.

%1
%\begin{thebibliography}{99}
%

%\end{thebibliography}

\newpage
\clearpage
\setcounter{figure}{0}

\end{document}